\documentclass[prd,aps,nofootinbib,
showkeys]
{revtex4}
\usepackage{graphicx,epsf,amsfonts,amssymb,amsbsy}
\newcommand{\ds}{\displaystyle}
\newcommand{\vev}[1]{\langle#1\rangle}
\newcommand{\mat}{\left ( \begin{array}}
\newcommand{\emat}{\end{array} \right )}
\newcommand{\vect}{\left ( \begin{array}{c}}
\newcommand{\evect}{\end{array} \right )}

\begin{document}

\title{ \bf Dual properties of dense quark matter with
color superconductivity phenomenon
}

\author{
T. G. Khunjua $^{1}$, K. G. Klimenko $^{2}$, and R. N. Zhokhov $^{2,3}$ }

\affiliation{$^{1}$ The University of Georgia, GE-0171 Tbilisi, Georgia}
\affiliation{$^{2}$ State Research Center
of Russian Federation -- Institute for High Energy Physics,
NRC "Kurchatov Institute", 142281 Protvino, Moscow Region, Russia}
\affiliation{$^{3}$  Pushkov Institute of Terrestrial Magnetism, Ionosphere and Radiowave Propagation (IZMIRAN),
108840 Troitsk, Moscow, Russia}

\begin{abstract}
In this paper the massless NJL model extended by the diquark interaction channel is considered. We study its phase 
structure at zero temperature and in the presence of baryon $\mu_B$, 
isospin $\mu_I$, chiral $\mu_{5}$ and chiral isospin $\mu_{I5}$  chemical potentials in the mean-field approximation.
It is shown that the model thermodynamic potential, which depends on three order parameters, $M$, $\pi_1$ and $\Delta$ 
(where $M$, $\pi_1$, and $\Delta$ -- are, respectively, chiral, charged pion and diquark
condensates of the model), is symmetric with respect to the (dual) transformation when $M\leftrightarrow\pi_1$ and 
simultaneously $\mu_I\leftrightarrow\mu_{I5}$. As a result, on the mean-field phase portrait of the model the chiral symmetry 
breaking (CSB) and charged pion condensation (PC) phases turn out to be dually conjugate with each other, which greatly 
simplifies the study of the phase portrait of the model. In particular, the duality between CSB and charged PC phases 
means that in the $(\mu_I,\mu_{I5})$-phase portrait these phases are mirror-symmetrical with respect to the line 
$\mu_I=\mu_{I5}$, which at the same time is the symmetry axis of the color superconducting (CSC) phase. Moreover, 
it follows from our analysis that chiral $\mu_5$ chemical potential promotes the formation of CSC phase in dense 
quark matter. And together with $\mu_{I5}$ it can generates the charged PC phase even at $\mu_I=0$.
\end{abstract}


\maketitle
\maketitle

\section{Introduction}

It is well known that quantum chromodynamics (QCD) is the theoretical basis for the study of strongly interacting 
matter. However, to consider the properties of dense quark (baryonic) matter, which can exist in the cores of compact 
stars or arise in collisions of heavy ions, perturbative QCD methods are inapplicable, since the coupling constant 
of strong interactions is quite large. In this case, effective QCD-like models, including the simplest Nambu--Jona-Lasinio (NJL) model 
which describes four-fermion interaction of $u$ and $d$ quarks \cite{njl,buballa,buballa2,zhokhov}, are often used to 
study dense baryonic medium composed of $u$ and $d$ quarks. 

Extending this model by the terms with baryon $\mu_B$ and isospin $\mu_I$ chemical potentials, one can investigate
the properties of quark matter with nonzero baryon $n_B$ and nonzero isospin $n_I$ densities (which is typical 
for neutron stars) \cite{son,he,ak,ekkz,Mammarella:2015pxa,Andersen:2018nzq}. More recently, it has become clear that 
chiral asymmetry (or chiral imbalance), i.e. unequal densities $n_L$ and $n_R$ of all left- and all right-handed
quarks, is 
also one of the properties of dense quark matter. Usually the chiral asymmetry is characterized by the quantity $n_5$,
called the chiral density, $n_5\equiv n_R-n_L$. It can be generated dynamically at high temperatures, such as in a 
fireball after a heavy ion collision, due to the Adler-Bell-Jackiw anomaly and the interaction of quarks with gauge 
(gluon) field configurations with non-trivial topology, called sphalerons. However, in the most general case the 
chiral densities of $u$ and $d$ quarks, i.e. the values $n_{u5}\equiv n_{uR}-n_{uL}$ and $n_{d5}\equiv n_{dR}-n_{dL}$,
are not equal and the quantity $n_{I5}\equiv n_{u5}-n_{d5}$ is also nonzero (it is the so-called chiral isospin 
density of quark matter). As it was discussed in Ref. \cite{Khunjua:2019lbv}, the nonzero $n_{I5}$ can be created in dense quark matter in the 
presence of rather strong magnetic fields and even not so large values of temperature, i.e. chiral isospin density can 
be observed in magnetars. As a result, we see that baryon medium composed of $u$ and $d$ quarks can be characterized,
in addition to temperature, by four more physical parameters, i.e. by densities $n_B$, $n_I$, $n_{5}$ 
and $n_{I5}$. Its thermodynamics can be studied, in particular, on the basis of the simplest NJL model extended with 
four chemical potentials $\mu_B$, $\mu_I$, $\mu_5$ and $\mu_{I5}$, which are thermodynamically conjugated to the 
corresponding densities. 

In the mean-field approximation, the $(\mu_B,\mu_I,\mu_5,\mu_{I5})$-phase portrait of this {\it massless} NJL model 
has been investigated in the papers \cite{kkz18,kkz18-2}, where it was shown that only two non-trivial phases are allowed 
for dense quark matter: (i) the chiral symmetry breaking (CSB) phase and (ii) the charged pion 
condensation (PC) one. Moreover, it was established that CSB and charged PC phases are dually conjugated to each 
other. It means that at fixed $\mu_B$ and $\mu_5$ these phases are arranged mirror-symmetrically 
with respect to the line $\mu_I=\mu_{I5}$ on the mean-field $(\mu_I,\mu_{I5})$-phase portrait of the model, and so on.
A more detailed influence of this dual symmetry on the phase structure of quark matter considered in the framework 
of the simplest massless NJL model 
has been investigated in Refs. \cite{zhokhov,kkz18,kkz18-2,khunjua,khunjua2}. As a result, it became clear
that duality greatly simplifies the study of the phase structure of quark matter, especially when it is under the 
influence of several external factors (chemical potentials). However, it must be borne in mind that the duality 
between CSB and charged PC is exact (in the large-$N_c$ or mean-field approximations)
only in the chiral limit, i.e. when the bare quark mass $m$ is zero. At $m\ne 0$ the duality between
CSB and charged PC is only approximative \cite{khunjua2}. Nevertheless, the NJL analysis at the physical point and 
observed approximative duality between these phases are in fairly good agreement with lattice QCD. 
In addition, it is important to be aware that duality has been proven only within the framework of the simplest 
NJL model, which is equivalent effectively to QCD only in the region of rather small baryon densities, i.e. at 
$\mu_B<1$ GeV. At higher energies or values of chemical potentials a more complicated NJL models, whose Lagrangians 
contain 
other quark interaction channels, should be taken into account. As a result, the set of possible ground states of 
dense quark matter can be expanded, which (in the absence of dual symmetry between some phases of matter) leads to 
significant difficulties in studying the phase structure of real quark matter, especially in the presence of four 
chemical potentials. The solution of the task could be greatly simplified if we had confidence in the existence of 
duality between some phases of matter even at rather large values of $\mu_B$, i.e. in the framework of a 
more complex NJL models, which have a rather diverse structure of four-quark interactions. 
The proposed work is devoted to the consideration of this problem.

An illustration of the above can be the situation with quark matter inside neutron stars, whose baryon density
is several times greater than the density of ordinary nuclear matter, i.e. in this case $\mu_B\gtrsim 1$ GeV. Moreover, 
there exist an evident isospin and chiral asymmetries. Under such extrim conditions, not only quark-antiquark pairing 
is possible, but also the formation of diquark pairs, as well as their condensation. It means that for large values 
of $\mu_B$ the phenomenon of color superconductivity (CSC) can be observed in the system. Therefore, in order to make
it more adequate an investigation of quark matter properties, an extension of the standard NJL model with the help of 
four-fermion terms responsible for the diquark interaction channel is usually used (see, e.g., the reviews 
\cite{buballa,alford} or Eq. (\ref{1}) below). The main goal of the present paper is to show that in this case, despite the 
complication of the model, its thermodynamic potential as a function of order parameters and four above mentioned 
chemical potentials has in the mean-field approximation a dual symmetry between the CSB and the charged PC phases. 
Thereby, the study of the phase structure of the model is simplified, and it is possible to shed new light on the 
role of both isospin and chiral (isospin) asymmetries in the formation of diquark condensation. Moreover, our results 
can serve as an argument in favor of the fact that duality is a characteristic feature of the QCD itself, and not
only of its different effective models.

The paper is organized as follows. In Sec. II a (3+1)-dimensional NJL model with two massless quark flavors ($u$ and 
$d$ quarks) that includes four kinds of chemical potentials, $\mu_B,\mu_I,\mu_{I5},\mu_{5}$, is introduced. 
In addition to usual quark-antiquark channels, the model contains the diquark interaction one and is intended to 
describe the phenomenon of color superconductivity in a dense quark medium. Furthermore, the symmetries of the model 
are discussed and its thermodynamic potential (TDP) is presented in the mean-field approximation.
In Sec. III the definition of the dual symmetry of the TDP is given. It means that TDP is invariant under some 
interchange of chemical potentials as well as, in some cases, simultaneous interchange of condensates.
In particular, it was established, and this is one of the main results of the paper, that in the mean-field 
approximation the TDP of the model is invariant under the following two simultaneous transformations, 
$\mu_I\leftrightarrow\mu_{I5}$ and chiral condensate$\leftrightarrow$charged PC one. As a result, in the 
different $(\mu_I,\mu_{I5})$-phase portraits of the model CSB and charged PC phases are dually conjugated to 
each other, i.e. they locate mirror symmetrically with respect to the line $\mu_I=\mu_{I5}$.
In Sec. IV summary and conclusions are given. Some technical details are relegated to Appendices A and B.

\section{The model and its thermodynamic potential}

Our investigation is based on the NJL type model with two quark
flavors. Its Lagrangian describes the interaction in the 
quark--antiquark as well as scalar diquark channels,
\begin{eqnarray}
 L=\bar q\Big [\gamma^\nu i\partial_\nu-
m\Big ]q+ G\Big [(\bar qq)^2+
(\bar qi\gamma^5\vec\tau q)^2\Big ]+H\sum_{A=2,5,7}
[\overline{ q^c}i\gamma^5\tau_2\lambda_{A}q]
[\bar qi\gamma^5\tau_2\lambda_{A} q^c],
\label{1}
\end{eqnarray}
where the quark field $q\equiv q_{i\alpha}$ is a flavor doublet
($i=1,2$; alternatively we use the notations $q_1=u$ and $q_2=d$) and color triplet ($\alpha=1,2,3$ or
$\alpha=r,g,b$) as well as a four-component Dirac spinor;
$q^c=C\bar q^T$ and $\overline{q^c}=q^T C$ are charge-conjugated
spinors, and $C=i\gamma^2\gamma^0$ is the charge conjugation
matrix (the symbol $T$ denotes the transposition operation). It is
supposed that $u$ and $d$ quarks have an equal current (bare) mass
$m$. Furthermore,  $\tau_a$ stands for Pauli matrices, and
$\lambda_A$ for Gell-Mann matrices in flavor and color space,
respectively. Clearly, the Lagrangian $L$ is invariant under
transformations from color SU(3)$_c$ as well as baryon U(1)$_B$
groups. In addition, at $m=0$ this Lagrangian is invariant under
the chiral SU(2)$_L\times$SU(2)$_R$ group. At $m\ne 0$ the chiral
symmetry is broken to the diagonal isospin subgroup  SU(2)$_I$
with the generators $I_k=\tau_k/2$ ($k=1,2,3$). Moreover, in our
system the electric and baryonic charges are conserved quantities, too, since $Q=I_3+B/2$,
where $I_3$ is the third generator of the isospin group SU(2)$_I$,
$Q$ is the electric charge generator, and $B$ is the baryon charge
generator (evidently,  these quantities are unit matrices in color
space, but in flavor space they are: $Q=diag(2/3,-1/3)$,
$I_3=diag(1/2,-1/2)$ and $B=diag(1/3,1/3)$). If the Lagrangian
(\ref{1}) is obtained from the QCD one-gluon exchange
approximation, then the quantity $H/G$ should not differ too much from the value of 0.75 \cite{buballa,alford}. 

Note that the Lagrangian (\ref{1}) describes physical processes in vacuum. In order to use the model (1) to study properties
of dense quark medium, it is necessary to modify Lagrangian (1) by adding terms with chemical potentials,
\begin{eqnarray}
L_{dense}=L+\bar q{\cal M}\gamma^0 q\equiv L+\bar q\left [\frac{\mu_B}{3}+\frac{\mu_I}2\tau_3+\frac{\mu_{I5}}2\gamma^5\tau_3+
\mu_5\gamma^5\right ]\gamma^0 q,
  \label{2}
\end{eqnarray}
where the chemical potential matrix ${\cal M}$ contains in the most general case four different chemical potential 
terms responsible for the description of quark matter with nonzero baryon ($\mu_B$), isospin ($\mu_I$), chiral ($\mu_5$) and chiral 
isospin ($\mu_{I5}$) densities, respectively.

If all chemical potentials in (\ref{2}) are nonzero 
quantities, then $SU(2)_I$ at $m\ne 0$ is not the
symmetry group of this Lagrangian. Instead, due to the $\mu_{I}$ term, it is symmetric under the flavor $U(1)_{I_3}$ group,
$q\to\exp (\mathrm{i}\alpha\tau_3/2) q$. Note however, that in the chiral limit ($m=0$) an additional symmetry of the Lagrangian (\ref{2}) 
appears, $U(1)_{AI_3}:q\to\exp (\mathrm{i}\alpha\gamma^5\tau_3)q$.

To study the phase diagram of the system (\ref{2}), we need to get its 
thermodynamic potential (in the mean-field approximation). For this purpose, let us consider the linearized version ${\cal L}$ of Lagrangian (\ref{2})
that contains auxiliary bosonic fields,
\begin{eqnarray}
{\cal L}\ds &=&\bar q\Big [\gamma^\nu i\partial_\nu
+{\cal M}\gamma^0
 -\sigma - m -i\gamma^5\vec\pi\vec\tau\Big ]q
 -\frac{1}{4G}\Big [\sigma^2+\pi_a^2\Big ]
 \nonumber\\ &-&\frac1{4H}\Delta^{*}_{A}\Delta_{A}-
 \frac{\Delta^{*}_{A}}{2}[\overline{q^c}i\gamma^5\tau_2\lambda_{A} q]
-\frac{\Delta_{A'}}{2}[\bar q i\gamma^5\tau_2\lambda_{A'}q^c].
\label{3}
\end{eqnarray}
In Eq. (\ref{3}) and later a summation over  repeated
indices $a=1,2,3$ and $A,A'=2,5,7$ is implied.
Clearly, the Lagrangians (\ref{2}) and (\ref{3}) are equivalent, as
can be seen by using the equations of motion for bosonic fields,
which take the form 
\begin{eqnarray}
\sigma (x)=-2G(\bar qq),~~\pi_a(x)=-2G(\bar qi\gamma^5\tau_a q),~~
\Delta_{A}(x)\!\!&=&\!\!-2H(\overline{q^c}i\gamma^5\tau_2\lambda_{A}q),~~
\Delta^{*}_{A}(x)=-2H(\bar qi\gamma^5\tau_2\lambda_{A} q^c).
\label{4}
\end{eqnarray}
It follows from (\ref{4}) that the mesonic fields
$\sigma(x),\pi_a(x)$ are real quantities, i.~e.\
$(\sigma(x))^\dagger=\sigma(x),~~
(\pi_a(x))^\dagger=\pi_a(x)$ (the superscript symbol $\dagger$
denotes the hermitian conjugation), but all diquark fields
$\Delta_{A}(x)$ are complex scalars, so
$(\Delta_{A}(x))^\dagger=\Delta^{*}_{A}(x)$.
Clearly, the real $\sigma(x)$ and $\pi_a(x)$ fields
are color singlets, whereas scalar diquarks
$\Delta_{A}(x)$ form a color antitriplet $\bar 3_c$ of the
SU(3)$_c$ group. Note that the auxiliary bosonic field $\pi_3(x)$ corresponds to real $\pi^0(x)$ meson, whereas the physical $\pi^\pm(x)$-meson
fields are the following combinations of the composite fields (\ref{4}), $\pi^\pm(x)=(\pi_1(x)\mp i\pi_2(x))/\sqrt{2}$. If some of the scalar 
diquark fields have a nonzero ground state expectation value, i.~e.\  $\vev{\Delta_{A}(x)}\ne 0$, 
the color symmetry of the model (\ref{2}) is spontaneously broken down.
 
In the one fermion-loop
approximation (or in the mean-field approximation) and  in the presence of dense quark medium, the effective action ${\cal S}_{\rm
{eff}}(\sigma,\pi_a,\Delta_{A},\Delta^{*}_{A'})$ of the model (1) is expressed by means of the path integral over quark
fields,
\begin{eqnarray}
\exp(i {\cal S}_{\rm {eff}}(\sigma,\pi_a,\Delta_{A},
\Delta^{*}_{A'}))=
  N'\int[d\bar q][dq]\exp\Bigl(i\int {\cal L}\,d^4 x\Bigr),\label{10}
\end{eqnarray}
where
\begin{eqnarray}
&&{\cal S}_{\rm {eff}}
(\sigma,\pi_a,\Delta_{A},\Delta^{*}_{A'})
=-\int d^4x\left [\frac{\sigma^2+\pi^2_a}{4G}+
\frac{\Delta_{A}\Delta^{*}_{A}}{4H}\right ]+
\tilde {\cal S}_{\rm {eff}},
\label{11}
\end{eqnarray}
and $N'$ is a normalization constant.
The quark contribution to the effective action, i.~e.\  the term
$ \tilde {\cal S}_{\rm {eff}}$ in (\ref{11}), is the following one, 
\begin{eqnarray}
\exp(i\tilde {\cal S}_{\rm {eff}})=N'\int [d\bar
q][dq]\exp\Bigl(i\int\Big [\bar q Dq-
 \frac{\Delta^{*}_{A}}{2}[\overline{q^c}i\gamma^5\tau_2\lambda_{A} q]
-\frac{\Delta_{A'}}{2}[\bar q i\gamma^5\tau_2\lambda_{A'}q^c]\Big ]d^4 x\Bigr),
\label{12}
\end{eqnarray}
where we have used the notation
\begin{eqnarray}
D=\big(\gamma^\nu i\partial_\nu
+{\cal M}\gamma^0
 -\sigma (x) - m -i\gamma^5\vec\pi (x)\vec\tau\big)\cdot 1\!\!{\rm I}_{3_c}, 
\label{13}
\end{eqnarray}
where $1\!\!{\rm I}_{3_c}$ is the unit operator in the tree-dimensional color space. Starting from Eqs. (\ref{11}) and (\ref{12}), one can define the thermodynamic potential (TDP) $\Omega(\sigma,\pi_a, \Delta_{A}, \Delta^{*}_{A'})$ of the model
(\ref{1}).  Indeed, this quantity is defined by the following relation,
\begin{equation}
{\cal S}_{\rm {eff}}~\bigg
|_{~\sigma,\pi_a,\Delta_{A},\Delta^{*}_{A'}=\rm {const}}
=-\Omega(\sigma,\pi_a,\Delta_{A},\Delta^{*}_{A'})\int d^4x.
\label{17}
\end{equation}
The ground state expectation values (mean
values) $\vev{\sigma(x)},~\vev{\pi_a(x)},~\vev {\Delta_{A}(x)},~\vev{\Delta^{*}_{A'}(x)}$ of the auxiliary bosonic fields (\ref{4}) are
solutions of the gap equations for the TDP $\Omega(\sigma,\pi_a,\Delta_{A},\Delta^{*}_{A'})$ (in our approach all ground state expectation values do not depend on
coordinates $x$), \footnote{In thermodynamics, the quantities $\vev{\sigma(x)},...$ are usually called order 
parameters, which determine, in essence, the phase structure of the system.}
\begin{eqnarray}
\frac{\partial\Omega}{\partial\pi_a}=0,~~~~~
\frac{\partial\Omega}{\partial\sigma}=0,~~~~~
\frac{\partial\Omega}{\partial\Delta_{A}}=0,~~~~~
\frac{\partial\Omega}{\partial\Delta^{*}_{A'}}=0,
\label{18}
\end{eqnarray}
and usually they are the coordinates of the global minimum point (GMP) of $\Omega$ vs. $\sigma,\pi_a,\Delta_{A},\Delta^{*}_{A'}$.
Note that at nonzero bare quark mass, $m\ne 0$, the quantity $\vev{\sigma(x)}$, called chiral condensate, is always nonzero. However, 
in the chiral limit, $m=0$, there can exists a region of chemical potentials in which $\vev{\sigma(x)}=0$. In this case it is the chiral symmetric 
phase of the model. In contrast, at $m=0$ the region with $\vev{\sigma(x)}\ne 0$ is called the chiral symmetry
breaking (CSB) phase. Moreover, both 
at $m=0$ and $m\ne 0$ the GMP of the TDP with $\vev{\pi^\pm(x)}\ne 0$ corresponds to the so-called charged pion condensation (PC) phase. 
But in the case when one of the scalar diquark fields has a nonzero ground state expectation value,  $\vev {\Delta_{A}(x)}\ne 0$, we have a phase with 
spontaneously broken color $SU(3)_c$ symmetry. It is called the color superconducting (CSC) one.

Recall that both the Lagrangian (\ref{2}) and the effective action (\ref{11}) are invariant under the color $SU(3)_c$ group. As a consequence of this fact, all $\Delta_A$ and 
$\Delta^*_{A'}$-dependence of the TDP $\Omega(\sigma,\pi_a,\Delta_{A},\Delta^{*}_{A'})$ (\ref{17}) is shown in its form depending on the combination $\Delta_2\Delta^*_2+\Delta_5\Delta^*_5
+\Delta_7\Delta^*_7\equiv \Delta^2$, where $\Delta$ is a real quantity. 
Let us note also that in the chiral limit (due to a $U_{I_3}(1)\times U_{AI_3}(1)$ invariance of the model (2)) 
the TDP $\Omega$ (\ref{17}) depends effectively only on the combinations $\sigma^2+\pi_3^2$ and $\pi_1^2+\pi_2^2$ 
(in addition to $\Delta$). So at $m=0$ one can put $\pi_3=0$ and $\pi_2=0$ without loss of generality.  Whereas at 
the physical point (i.e. at $m\ne 0$) it depends effectively on the combination $\pi_1^2+\pi_2^2$ as well as on 
$\sigma$ and $\pi_3$. Since in this case the relations $\vev{\sigma(x)}\ne 0$ and $\vev{\pi_3(x)} = 0$ are always 
satisfied in the NJL model (2) at $H=0$ \footnote{See, e.g., the gap equations (13) and (14) of Ref. \cite{Ebert}, which 
at $m\ne 0$ have single solution $\vev{\sigma(x)}\ne 0$ and $\vev{\pi_3(x)} = 0$.}, at $m\ne 0$ one can made a rather plausible assumption that in Eq. (\ref{17}) $\pi_2=\pi_3=0$ is also valid 
and study this TDP as a function of only three variables, $\sigma$, $\pi_1$ and $\Delta$, i.e. $\Omega\equiv\Omega(\sigma,\pi_1,\Delta)$. 
It is clear that in order to calculate $\Omega(\sigma,\pi_1,\Delta)$ it is enough to suppose 
that in Eqs. (\ref{11}) and (\ref{12}) $\Delta_2=\Delta_2^*=\Delta$, $\Delta_5=\Delta_7=0$ and $\pi_2=\pi_3=0$ 
(note that in the following we also suppose that all auxiliary bosonic fields do not depend on space coordinate $x$).  
Since
\begin{eqnarray*}
\lambda_2=\left (\begin{array}{ccc}
0, & \!\!\!-i,&0\\
i~, & 0,&0\\
0,&0,&0
\end{array}\right )=\left (\begin{array}{cc}
\sigma_2, & 0\\
0~,&0 
\end{array}\right ),
\end{eqnarray*}
where $\sigma_2$ is the corresponding Pauli matrix acting in the two-dimensional fundamental representation of the $SU(2)_c$ subgroup 
of the $SU(3)_c$, 
it is clear that uder this assumption the contribution of the blue $q_b$ quarks in the expression  (\ref{12}) is factorized. Then
\begin{eqnarray}
\exp(i\tilde {\cal S}_{\rm {eff}})&=&N'\int [d\bar
q_b][dq_b]\exp\Bigl(i\int\Big [\bar q_b D^+q_b\Big ]\Bigr)\nonumber\\
&\times&
\int [d\overline Q][dQ]\exp\Bigl(i\int\Big [\overline Q \big(D^+\cdot 1\!\!{\rm I}_{2_c}\big) Q-
 \frac{\Delta}{2}[\overline{Q^c}i\gamma^5\tau_2\sigma_{2} Q]
-\frac{\Delta}{2}[\overline Q i\gamma^5\tau_2\sigma_{2}Q^c]\Big ]d^4 x\Bigr),
\label{21}
\end{eqnarray}
where $q_b$ is the flavor doublet of the blue quarks and $Q$ is the flavor ($u$ and $d$) and color (red and green) quark doublet. Moreover, here 
we use the notation $D^+$ for the operator that is in the round brackets of Eq. (\ref{13}) at $\pi_2=\pi_3=0$ (see below in Eq. (\ref{14})) 
and $1\!\!{\rm I}_{2_c}$ is 
the unit operator in the two-dimensional color space.
Performing in Eq. (\ref{21}) the functional integrations over $q_b$ (which is a trivial one) and over $Q$ (see Appendix \ref{AA}), we find
\begin{eqnarray}
\exp(i\tilde {\cal S}_{\rm {eff}})&=&N'\det D^+\cdot{\det}^{1/2}(Z),
\label{22}
\end{eqnarray}
where
\begin{equation}
Z=\left (\begin{array}{cc}
D^+\cdot 1\!\!{\rm I}_{2_c}, & -K\\
~-K~~~~ , &D^-\cdot 1\!\!{\rm I}_{2_c}
\end{array}\right ),\label{15}
\end{equation}
and
 \begin{eqnarray}
&&D^+=i\gamma^\nu\partial_\nu- m+{\cal M}\gamma^0-\Sigma,~~~\Sigma=\sigma+ i\gamma^5\pi_1\tau_1,
\nonumber\\
&&D^-=i\gamma^\nu\partial_\nu- m-\gamma^0 {\cal M}-\Sigma,~~~
K=i\Delta\gamma^5\tau_2\sigma_{2}.
\label{14}
\end{eqnarray}
Note that matrix elements of the matrix $Z$ (\ref{15}) are the operators in two-dimensional color and flavor spaces as well as in four-dimensional spinor and 
coordinate spaces. Then, it follows from Eqs. (\ref{11}) and (\ref{22}) that 
\begin{equation}
{\cal S}_{\rm
{eff}}(M,\pi_1,\Delta)
=-\int d^4x\left[\frac{(M-m)^2+\pi^2_1}{4G}+
\frac{\Delta^2}{4H}\right]-\frac i2\ln\det (Z)-i\ln\det (D^+),
\label{16}
\end{equation}
where we have introduced the gap $M\equiv\sigma+m$. The last term of Eq. (\ref{16}), which does not depend on $\Delta$, was calculated 
in our recent paper \cite{kkz18} (see there Eqs. (16)-(21)),
\begin{eqnarray}
i\ln\det (D^+)=i\int\frac{d^4p}{(2\pi)^4}\ln\Big [\big (\eta^4-2a_+\eta^2+b_+\eta+c_+\big )\big (\eta^4-2a_-\eta^2+b_-\eta+c_-\big )\Big ]
\int d^4x,
\label{07}
\end{eqnarray}
where $\eta=p_0+\mu$, $|\vec p|=\sqrt{p_1^2+p_2^2+p_3^2}$ and
\begin{eqnarray}
a_\pm&=&M^2+\pi_1^2+(|\vec p|\pm\mu_{5})^2+\nu^2+\nu_{5}^2;~~b_\pm=\pm 8(|\vec p|\pm\mu_{5})\nu\nu_{5};\nonumber\\
c_\pm&=&a_\pm^2-4 \nu ^2
\left(M^2+(|\vec p|\pm\mu_{5})^2\right)-4 \nu_{5}^2 \left(\pi_1^2+(|\vec p|\pm\mu_{5})^2\right)-4\nu^{2} \nu_{5}^2
\label{101}
\end{eqnarray}
(we also use in Eq. (\ref{101}) and below the notations $\mu=\mu_B/3$, $\nu=\mu_I/2$ and $\nu_5=\mu_{I5}/2$). The 
next undefined term of Eq. (\ref{16}) is the following
\begin{eqnarray}
\frac i2\ln\det (Z)=i\int\frac{d^4p}{(2\pi)^4}\ln\det L(p)\int d^4x,
\label{20}
\end{eqnarray}
where $L(p)$ is given by Eq. (\ref{IV.34}), i.e. it is the 4$\times$4 matrix in spinor space. 
Taking into account the relations (\ref{16})-(\ref{20}) as well as the definition (\ref{17}), 
it is easy to get the TDP of the model in the mean-field approximation, 
\begin{eqnarray}
\Omega(M,\pi_1,\Delta)
&=&\left[\frac{(M-m)^2+\pi^2_1}{4G}+
\frac{\Delta^2}{4H}\right]+i\int\frac{d^4p}{(2\pi)^4}\ln\det L(p)\nonumber\\
&+&i\int\frac{d^4p}{(2\pi)^4}\ln\Big [\big (\eta^4-2a_+\eta^2+b_+\eta+c_+\big )\big (\eta^4-2a_-\eta^2+b_-\eta+c_-\big )\Big ].
\label{19}
\end{eqnarray}
In what follows, when performing numerical calculations or finding symmetry properties of the TDP 
$\Omega(M,\pi_1,\Delta)$, it is very convenient to represent $\det L(p)$ in Eq. (\ref{19}) as a production of 
four eigenvalues of the matrix $L(p)$,
\begin{eqnarray}
\det L(p)=\tilde\lambda_1(p)\tilde\lambda_2(p)\tilde\lambda_3(p)\tilde\lambda_4(p).
\label{21}
\end{eqnarray}
Here
\begin{eqnarray}
&&\widetilde\lambda_{1,2}(p)=\lambda_{1,2}(p)\Big |_{|\vec p|\to|\vec p|-\mu_5},~~\widetilde\lambda_{3,4}(p)=
\lambda_{3,4}(p)\Big |_{|\vec p|\to|\vec p|+\mu_5}
\label{71}
\end{eqnarray}
and
\begin{eqnarray}
&&\lambda_{1,2}(p)=N_1\pm 4\sqrt{K_1},~~\lambda_{3,4}(p)=N_2\pm 4\sqrt{K_2},
\label{56}
\end{eqnarray}
where
\begin{eqnarray}
\hspace{-1cm}N_2=N_1+16\mu\nu\nu_5|\vec p|,~~K_2=K_1+8\mu\nu\nu_5|\vec p|p_0^4-8\mu\nu\nu_5|\vec p|p_0^2\big (M^2+\pi_1^2+|\Delta|^2
+|\vec p|^2+\mu^2+\nu^2-\nu_5^2\big ),
\label{57}
\end{eqnarray}
\begin{eqnarray}
K_1=\nu_5^2p_0^6-p_0^4\Big [2\nu_5^2 \big(|\Delta|^2+\pi_1^2+M^2+|\vec p|^2+\nu^2+\mu^2-\nu_5^2\big)+4\mu\nu\nu_5|\vec p|\Big ]+
p_0^2 \Big\{
\nu_5^6+2\nu_5^4 \big (M^2-|\Delta|^2-\pi_1^2&&~~~~~~~~~~~~~~\nonumber\\
-\nu^2-\mu^2-|\vec p|^2\big )+4\mu^2\nu^2\big(M^2+
|\vec p|^2\big)+4|\vec p|\mu\nu\nu_5\big(|\Delta|^2+\pi_1^2+M^2+|\vec p|^2+\nu^2+\mu^2-\nu_5^2\big)&&\nonumber\\
+\nu_5^2\Big [\big(|\Delta|^2+\pi_1^2+|\vec p|^2+\nu^2+\mu^2\big)^2+2|\vec p|^2M^2+M^4+2M^2\big(|\Delta|^2-\nu^2+\pi_1^2-\mu^2\big)\Big ]\Big\},&&
\label{58}
\end{eqnarray}
\begin{eqnarray}
N_1&=&p_0^4-2p_0^2\Big [|\Delta|^2+\pi_1^2+M^2+|\vec p|^2+\nu^2+\mu^2-3\nu_5^2\Big ]+
\nu_5^4-2\nu_5^2\Big [|\Delta|^2+\pi_1^2+|\vec p|^2+\nu^2+\mu^2-M^2\Big ]\nonumber\\
&-&8\mu\nu\nu_5|\vec p|+\left (|\vec p|^2+M^2+\pi_1^2+|\Delta|^2-\mu^2-\nu^2\right )^2-
4\left (\mu^2\nu^2-\pi_1^2\nu^2-|\Delta|^2\mu^2\right ).
\label{59}
\end{eqnarray}
Exact expressions for the eigenvalues $\tilde\lambda_i(p)$ were obtained earlier in the study of the phase 
structure of the two-color QCD (see, e.g., the section III of Ref. \cite{2color}). 
We are now ready to discuss in the chiral 
limit, i.e. at $m=0$, the dual properties of the NJL model (\ref{1}) in the mean-field approximation.

Throughout the paper we use in numerical investigations of the TDP (\ref{19}) 
the soft cut-off 
regularization scheme when $d^4p\equiv dp_0d^3\vec p\to dp_0d^3\vec p f_{\Lambda}(\vec p)$. 
Here the cut-off function is 
\begin{eqnarray}
f_{\Lambda}(\vec p)=\sqrt{\frac{\Lambda^{2N}}{\Lambda^{2N}+|\vec p|^{2N}}}, 
\end{eqnarray}
and the parameter fit used is $G=4.79$ GeV$^{-2}$, $\Lambda=638.8$ MeV and $N=5$.

\section{Dual symmetries of the TDP (\ref{19}) and phase portraits of the model}

First of all, we note once again that in order to find the phase structure of the NJL model (\ref{2}), it is 
necessary to study its TDP (\ref{19}) as a function of three variables $M,\pi_1,\Delta$ for an absolute minimum, and 
then see how the properties of its global minimum point (GMP) change depending on the values of chemical potentials
$\mu,\nu,\nu_{I5},\mu_5$. 

In this regard, we remark that in the chiral 
limit, $m=0$, for sufficiently low values of the chemical potentials (say at $\mu,\nu,\nu_5,\mu_5 <1$ GeV) at the 
GMP $(M,\pi_1,\Delta)$ of the TDP (\ref{19}) there can be no more than one nonzero coordinates. (This conclusion can 
be done based on the arguments presented earlier in our papers \cite{2color,kkz18}.) Therefore, with such 
a restriction on chemical potentials, in the chiral limit only four different phases can be realized in the system.
(I) If GMP has the form $(M\ne 0,\pi_1=0,\Delta=0)$, then the chiral symmetry breaking (CSB) phase appears in 
the model. (II) If it has the form $(M=0,\pi_1\ne 0,\Delta=0)$, the charged pion condensation (PC) phase is realized.
(III) When the GMP looks like $(M=0,\pi_1=0,\Delta\ne 0)$, it corresponds to the color superconducting (CSC) or
diquark condensation phase. And finally, (IV) the GMP of the form $(M=0,\pi_1=0,\Delta=0)$ corresponds to a 
symmetrical phase with all zero condensates.  
Then a phase portrait of the model is no more than a one-to-one correspondence between any point
$(\mu,\nu,\nu_5,\mu_5)$ of the four-dimensional space of chemical potentials and possible model
phases (CSB, charged PC, CSC and symmetric phase). However, it is clear that this four-dimensional
phase portrait (diagram) is quite bulky and it is rather hard to imagine it as a whole. So in order
to obtain a more deep understanding of the phase diagram as well as to get a greater visibility
of it, it is very convenient to consider different low-dimensional cross-sections of this general
$(\mu,\mu_{5},\nu,\nu_{5})$-phase portrait, defined by the constraints of the form $\mu= const$ and
$\mu_5=const$, etc. 

In addition, note that the study of the phase structure of any model is greatly simplified if the so-called dual symmetries of its TDP 
is taken into account. Recall that by dual symmetry of the TDP we understand its symmetry (invariance) with respect
to any discrete transformations as order parameters (in our case, these are $M$, $\pi_1$ and $\Delta$ condensates) 
and free external parameters of the system (these may be chemical potentials, coupling constants, etc).
The presence of the dual symmetry of the model TDP means that in its phase portrait there is some
symmetry between phases with respect to the transformation of external parameters. And this circumstance 
can simplify the construction of the full phase diagram of the system. 

For example, taking into account the Eqs. (\ref{101})-(\ref{21}), it is possible to establish the invariance of the 
TDP (\ref{19}) with respect to each of the following six transformations, in each of them two
chemical potentials change their sign simultaneously: (i) $\{\nu\to-\nu;~\nu_5\to-\nu_5\}$, (ii) $\{\nu\to-\nu;~\mu_5\to-\mu_5\}$, (iii)
$\{\nu_5\to-\nu_5;~\mu_5\to-\mu_5\}$, (iv) $\{\mu\to-\mu;~\mu_5\to-\mu_5\}$, (v) $\{\mu\to-\mu;~\nu\to-\nu\}$, and (vi)
$\{\mu\to-\mu;~\nu_5\to-\nu_5\}$. The invariance of the TDP (\ref{19}) under the transformations (i)--(vi) 
is one of the simplest examples of its dual symmetries that can help us to simplify the analysis of the phase portrait
of the model. In particular, due to the symmetry of the TDP (\ref{19}) under transformations (i)-(vi), it is 
sufficient to study the phase structure of the NJL model (\ref{2})
only, e.g.,  in the case when arbitrary three of the four chemical potentials have positive signs, whereas the sign of the rest chemical
potential is not fixed. Then, applying to a phase diagram with this particular distribution of the chemical potential signs one or several
transformations (i)--(vi), it is possible to find a phase portrait of the model at an arbitrary distribution of chemical potential signs.
Hence, by agreement, in the following we will study the phase diagram of the model (\ref{2}) only at $\mu\ge 0$, $\nu\ge 0$, $\nu_5\ge 0$ and for arbitrary sign of $\mu_5$.
\begin{figure}
\includegraphics[width=0.45\textwidth]{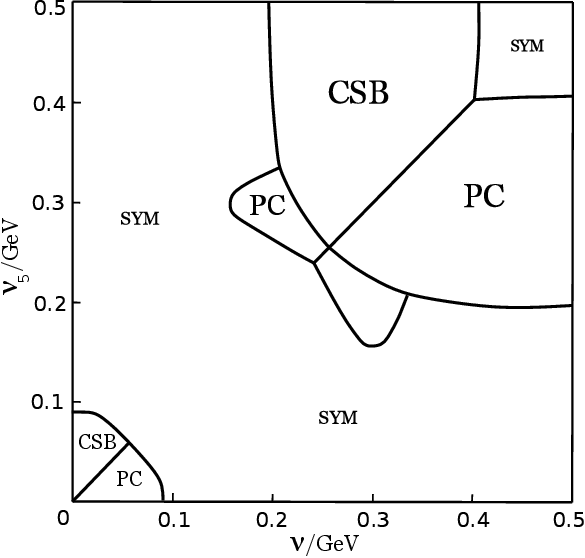}
 \hfill
\includegraphics[width=0.45\textwidth]{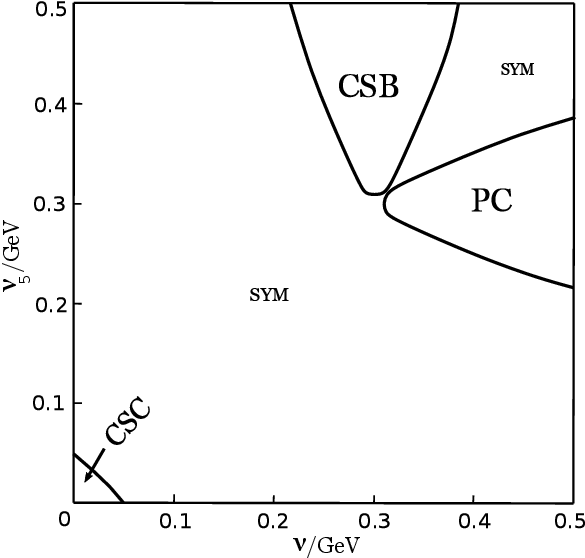}\\
\caption{Some $(\nu,\nu_{5})$-phase diagrams of the model at $H=0.75G$. Left panel: The case 
of $\mu=0.3$ GeV and $\mu_5=0$. 
Right panel: The case of $\mu=0.3$ GeV and $\mu_5=0.15$ GeV. Here PC denotes
the charged pion condensation phase, CSB and CSC mean respectively the chiral symmetry breaking and color 
superconducting phases, ``sym`` is the symmetric phase. 
 } 
\end{figure}
\begin{figure}
\includegraphics[width=0.45\textwidth]{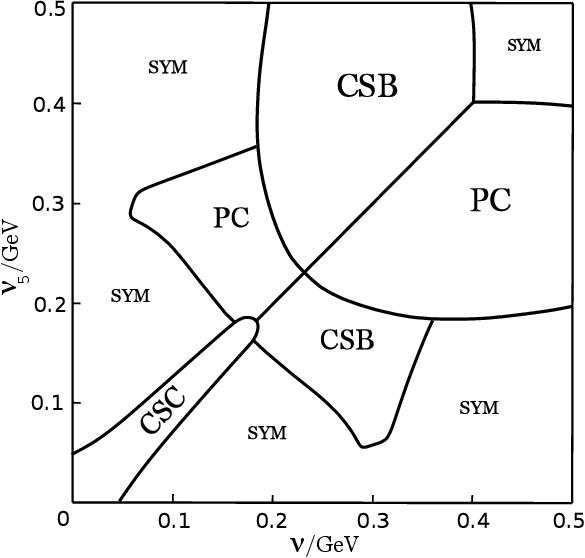}
 \hfill
\includegraphics[width=0.45\textwidth]{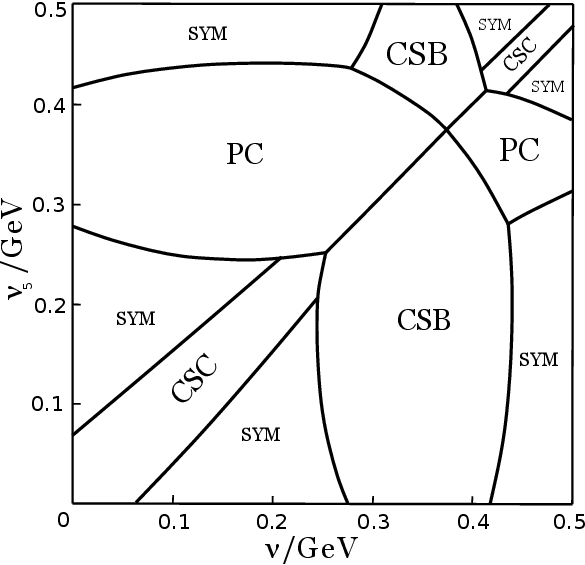}\\
\caption{ Some $(\nu,\nu_{5})$-phase diagrams of the model at $H=0.75G$. Left panel: The case of
$\mu=0.3$ GeV and $\mu_5=-0.1$ GeV. 
Right panel: The case of $\mu=0.3$ GeV and $\mu_5=-0.3$ GeV. All the notations are the same as in Fig. 1.
 } 
\end{figure}
\begin{figure}
\includegraphics[width=0.45\textwidth]{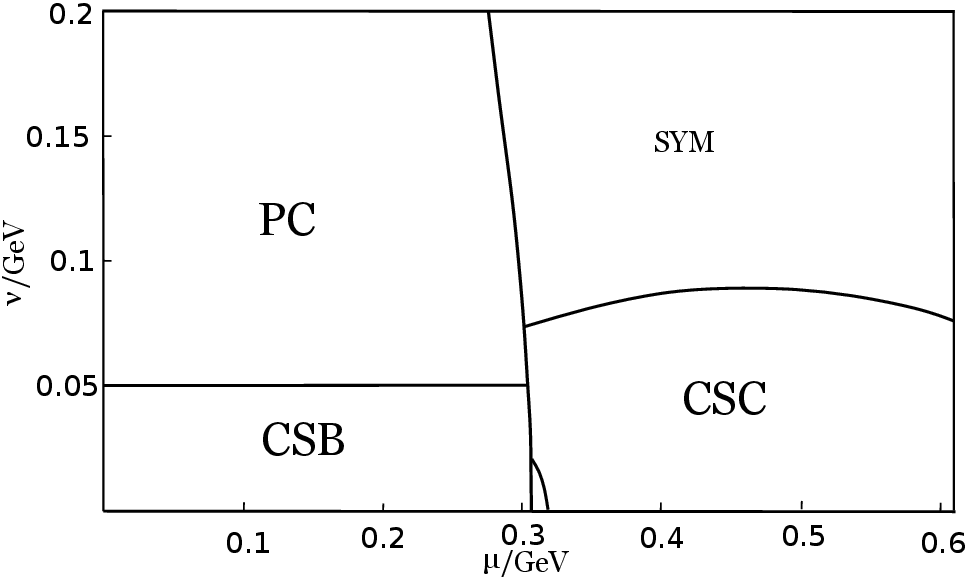}
 \hfill
\includegraphics[width=0.45\textwidth]{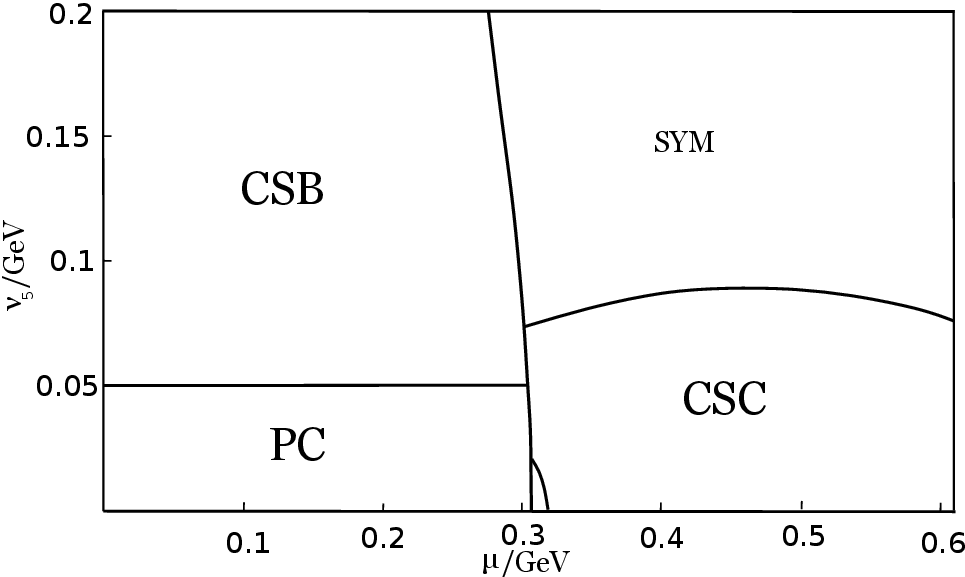}\\
\caption{  Left panel: $(\mu,\nu)$-phase portrait at $H=0.75G$. The case $\mu_5=-0.05$ GeV and $\nu_5=0.05$ GeV.  
Right panel: $(\mu,\nu_5)$-phase portrait at $H=0.75G$. The case $\mu_5=-0.05$ GeV and $\nu=0.05$ GeV.
All the notations are the same as in Fig. 1.
 } 
\end{figure}

Another and more non-trivial example of the dual symmetry of the TDP $\Omega(M,\pi_1,\Delta)$ (\ref{19}) is its 
invariance under the two discrete transformations ${\cal D}$,
\begin{eqnarray}
{\cal D}:
~~\nu\longleftrightarrow\nu_5,~~M\longleftrightarrow \pi_1,
\label{60}
\end{eqnarray}
in the chiral limit. The invariance of the last term of Eq. (\ref{19}) under the transformation ${\cal D}$ follows directly from Eqs. 
(\ref{07})-(\ref{101}). However, checking the ${\cal D}$-invariance of its $\det L(p)$ term (\ref{21}) is not such a 
simple task. However, this is true, since this fact is easy to establish taking into account the relations 
(\ref{71})-(\ref{59}) and using any program of analytical calculations.

To understand the physical meaning of the dual symmetry (\ref{60}), let us suppose that $m=0$ and that at 
the point $(\mu=a, \nu=b, \nu_5=c,\mu_5=d)$ of the mean-field phase portrait 
the GMP of the TDP (\ref{19}) lies, e.g., at the point of the condensate space of the form
$(M=A,\pi_1=B,\Delta=C)$. Then, due to a dual symmetry (\ref{60}) of the TDP, at the dually ${\cal D}$-conjugated 
point of the phase portrait, i.e. at the point $(\mu=a, \nu=c, \nu_5=b,\mu_5=d)$, the dually ${\cal D}$-conjugated 
phase should be located. Its condensate structure has the form $(M=B,\pi_1=A,\Delta=C)$. 
Hence, we see that if, e.g., $M=A=0,\pi_1=B\ne 0,\Delta=C=0$, then CSB phase corresponding to a GMP of the form
$M=B\ne 0,\pi_1=A=0,\Delta=C=0$ should be dually conjugated to the initial charged PC phase, and vice versa. 
But if at the original point there is symmetric or CSC phase, then it is also realized at the dually conjugated point.
Thus, knowing the phase of the model, which is
realized at some point of its $(\mu,\nu,\nu_5,\mu_5)$-phase portrait, we can predict which phases are arranged at the
dually conjugated points of this phase diagram. Moreover, the order parameter of the initial CSB phase
of the point $(\mu=a, \nu=b, \nu_5=c,\mu_5=d)$, i.e. the quantity $M=A$, is equal to the order
parameter $\pi_1=A$ of the ${\cal D}$-dually conjugated charged PC phase of the point
$(\mu=a, \nu=c, \nu_5=b,\mu_5=d)$ of the model phase portrait, etc.

As a consequence, we see that at some fixed values of $\mu$ and $\mu_5$, in the $(\nu,\nu_5)$-phase 
portrait of the model the CSB and charged PC phases should 
be mirror symmetrical to each other with respect to the $\nu=\nu_5$ line. And each of the remaining phases, i.e. 
symmetric or CSC one, must occupy an area symmetrical about this line. This fact is well illustrated by four 
$(\nu,\nu_5)$-phase portraits depicted in Figs. 1 and 2 at $\mu=0.3$ GeV and different values of $\mu_5$. 
Moreover, it is clear from Fig. 1 (left panel) that at 
$\mu_5=0$ and at rather small values of $\nu$ and $\nu_5$ the CSC phase is absent. But when $\mu_5$ is nonzero, this phase appears in a small neighborhood of 
the point $\nu=0$, $\nu_5=0$ (see other phase diagrams of Figs. 1, 2). 
Thus, chiral $\mu_5$ chemical potential
promotes the formation of 
the CSC phase in a dense baryonic medium. In addition, the phase diagram of Fig. 2 (right panel) confirms one more interesting 
property of $\mu_5$ (as well as of $\nu_5$ chemical potential) noted earlier in the framework of the ordinary 
NJL model \cite{kkz18} (see there Figs. 10, 11):
If $\mu_5$ is nonzero, then chiral isospin $\nu_{5}$ chemical potential generates charged pion condensation in 
dense quark matter (it is the PC phase in the right panel of Fig. 2) even if isospin $\nu$ chemical potential equals to zero 
(see also Figs. 10, 11 of Ref. \cite{kkz18}). We emphasize once again that in order for this generation to take place
one needs to have nonzero chiral chemical potential $\mu_{5}$. In contrast, as it was
discussed in \cite{kkz18-2}, this generation requires nonzero values of $\nu$ in the case of $\mu_{5}=0$. 
Hence, in the $\nu=0$ case chiral $\mu_{5}$  chemical potential can take the role of $\nu$ and allow this 
generation to happen. 

Finally, we would like to draw attention to the fact that at $m=0$ the duality transformation ${\cal D}$
(\ref{60}) of the TDP (\ref{19}) can also be applied to a mean-field phase portrait of the model as a whole. 
Namely, in this case, i.e. when acting on the phase diagram, it is necessary to
rename both the diagram axes and phases in such a way, that $\nu\leftrightarrow\nu_5$
and CSB$\leftrightarrow$charged PC. At the same time the $\mu$- and $\mu_5$-axes and CSC and
symmetrical phases should not change their names and positions. It is evident that after such ${\cal D}$
transformation the full $(\mu,\nu,\nu_5,\mu_5)$-phase diagram is mapped to itself, i.e. the most
general $(\mu,\nu,\nu_5,\mu_5)$-phase portrait of the model is self-${\cal D}$-dual. 
In a similar way it is clear that various $(\nu,\nu_{5})$-phase diagrams at fixed $\mu$ and $\mu_5$ values 
are transformed into themselves after applying to them the dual operation (\ref{60}), i.e. they  
are also self-${\cal D}$-dual (see, e.g., in Figs. 1, 2). But other cross-sections of the
full mean-field $(\mu,\nu,\nu_5,\mu_5)$-phase diagram, e.g., the $(\mu,\nu)$-phase portrait at some fixed
values of $\nu_5$ and $\mu_5$, are not invariant, in general, under the action of dual transformation ${\cal D}$.
As a result, a completely different phase portrait can be obtained. Hence, based on this mechanism, 
it is possible, having a well-known cross-section of the full phase
diagram of the model, to obtain its phase portrait in a less studied range of values of chemical
potentials. For example, In Fig. 3 (left panel) one can see the $(\mu,\nu)$-phase portrait at fixed $\nu_5=0.05$ GeV 
and $\mu_5=-0.05$ GeV. Applying to it the dual operation  ${\cal D}$ according to the rule described above, 
one can obtain (without any numerical calculations) its dual conjugation, i.e. the $(\mu,\nu_5)$-phase portrait 
at fixed $\nu=0.05$ GeV and $\mu_5=-0.05$ GeV (see the right panel of Fig. 3).

\section{Summary and conclusions}

In this paper the phase structure of the generalized massless NJL model (\ref{1}), which describes interactions 
both in quark-antiquark and diquark channels, is discussed at zero temperature and in the presence of four chemical  
potentials, baryon $\mu_B=3\mu$, isospin $\mu_I=2\nu$, chiral isospin $\mu_{I5}=2\nu_5$, and chiral $\mu_5$ chemical 
potentials, in the mean-field approximation. The model is intended to be a theoretical basis for considering the 
properties of dense quark matter, the ground state of which can be realized as one of the following phases:
CSB, charged PC, CSC, and symmetric phase. To find out which of the phases is implemented in the model, we have 
considered its thermodynamic potential in the mean field approximation (\ref{19}). But, it is obvious that even in 
this approximation, 
the study of this TDP to an absolute minimum is a rather difficult problem, especially in the presence of four
above mentioned chemical potentials. (By the way, we note that these chemical potentials are thermodynamically 
conjugated with the real physical characteristics of dense quark matter that can exist in the cores of neutron 
stars (see the Introduction).)

Previously, a similar problem arose when studying the phase structure of the simplest NJL model 
(without taking into account the diquark interaction channel) with the same four chemical potentials 
\cite{kkz18,kkz18-2}. However, it turns out that in the latter case, the TDP of the model in the mean-field 
approximation has (i) a more simple form since it is a function of only two order parameters, and (ii) it is 
invariant under the so-called dual transformation ${\cal D}$ (its form is given by Eq. (\ref{60})). Due to these 
circumstances, and in particular the property (ii), the solution of the problem is greatly simplified. 

In the present work, we show that in the more complex four-fermion NJL model (\ref{1}) the thermodynamic potential 
in the chiral limit and in the mean-field approximation also has dual discrete symmetry ${\cal D}$ (\ref{60}), 
which greatly simplifies the study of the phase structure of the model. And it is the main result of the paper. 
This dual symmetry between the CSB and 
the charged PC phases is especially clearly manifested in the so-called $(\nu,\nu_5)$-phase diagrams by the 
mirror-symmetric arrangement of these phases relative to the line $\nu=\nu_5$. At the same time, the symmetric and 
CSC phases in these diagrams are symmetrical with respect to the same line (see Figs. 1 and 2). In addition, acting 
by the dual transformation ${\cal D}$ on more well-known phase portraits, one can get an idea of the phase structure 
of the model in the region of less studied values of chemical potentials (compare the two diagrams in Fig. 3).

Hence, we see that in two qualitatively different QCD-like NJL models there is a duality between CSB and charged PC 
phases in the mean-field approximation. The conclusion suggests itself that the dual symmetry ${\cal D}$ is inherent 
in these models not only in the mean-field approximation, but is also a characteristic property of both their 
microscopic Lagrangians and their full thermodynamic potentials. Moreover, we believe that the full massless two-flavor QCD Lagrangian also has a dual symmetry between 
CSB and charged PC phenomena. In the future, these issues will be discussed in more detail.

We hope that our results might shed some new light on phase structure of dense quark matter with isospin
and chiral (isospin) imbalances and hence could be important for describing physics, for example, 
in an interior of the compact stars.

\appendix
\section{Derivation of expressions (\ref{22}) and (\ref{20})}
\label{AA}
Here we perform a functional integration over flavor and color quark-doublet fields $Q$ in Eq. (\ref{21}). 
First, let us consider the half of the quantity $\overline Q\big(D^+\cdot 1\!\!{\rm I}_{2_c}\big)Q$,
\begin{eqnarray}
\frac 12 \overline Q\big(D^+\cdot 1\!\!{\rm I}_{2_c}\big)Q =\frac 12
\overline Q \Big [\big(i\hat\partial-m+{\cal M}\gamma^0 -\sigma -i\gamma^5\tau_1\pi_1\big)\cdot 1\!\!{\rm I}_{2_c}\Big ]Q,\label{IV.6}
\end{eqnarray}
and try to rewrite it in terms of charge conjugated quark fields $Q^c$ and $\overline {Q^c}$ using the well-known relations, $\overline Q=(Q^c)^T C$ and 
$Q=C(\overline{Q^c})^T$. Now let us
apply to the expression (\ref{IV.6})
the well-known linear algebra relation, $x_\alpha O_{\alpha\beta}y_\beta=\pm y_\beta O^T_{\beta\alpha}x_\alpha$, where the sign $+(-)$
corresponds to the case when the quantities $x_\alpha$ and $y_\beta$ are commute (anticommute). Therefore, since $Q_c$ and $\overline {Q^c}$ are anticommuting fields,
it can be transformed to the following one
\begin{eqnarray}
\frac 12 \overline Q\big(D^+\cdot 1\!\!{\rm I}_{2_c}\big)Q&=&-\frac 12 \overline{Q^c}
\Big\{ CD^+C\Big\}^T\cdot 1\!\!{\rm I}_{2_c}Q^c=
\frac 12 \overline{Q^c}
\Big\{ CD^+C^{-1}\Big\}^T\cdot 1\!\!{\rm I}_{2_c}Q^c\nonumber\\&=&\frac 12 \overline{Q^c}
\Big\{-i(\gamma^\mu)^T\partial_\mu-m-C{\cal M}C^{-1}(\gamma^0)^T -\sigma -i\gamma^5\vec\tau_1\pi_1\Big\}^T\cdot 1\!\!{\rm I}_{2_c}Q^c\nonumber\\&=&
\frac 12 \overline{Q^c}
\Big\{i\gamma^\mu\partial_\mu-m-\gamma^0{\cal M} -\sigma -i\gamma^5\tau_1\pi_1\Big\}\cdot 1\!\!{\rm I}_{2_c}Q^c\equiv\frac 12 \overline{Q^c}
\big(D^-\cdot 1\!\!{\rm I}_{2_c}\big)Q^c.\label{IV.7}
\end{eqnarray}
It is clear from Eq. (\ref{IV.7}) that 
\begin{eqnarray}
D^-\equiv i\gamma^\mu\partial_\mu-m-\gamma^0{\cal M} -\sigma -i\gamma^5\tau_1\pi_1=\Big\{ CD^+C^{-1}\Big\}^T.\label{IV.70}
\end{eqnarray}
Note also that the transpose operation presented in Eq. (\ref{IV.7}) means both transposition of matrices in color, flavor and spinor spaces 
and transposition of the differentiation operator, $\partial_\nu^T=-\partial_\nu$. Moreover, we also have used there the following relations:
$C^{-1}=C^T=-C$, $C\gamma^\nu C^{-1}=-(\gamma^\nu)^T$,
$C\gamma^5C^{-1}=(\gamma^5)^T=\gamma^5$, $C{\cal M}C^{-1}={\cal M}$ and ${\cal M}^T={\cal M}$. Another half of the quantity 
$\overline Q\big(D^+\cdot 1\!\!{\rm I}_{2_c}\big)Q$ we still consider unchanged. So
\begin{eqnarray}
\overline Q\big(D^+\cdot 1\!\!{\rm I}_{2_c}\big)Q=\frac 12\overline Q\big(D^+\cdot 1\!\!{\rm I}_{2_c}\big)Q+\frac 12 \overline{Q^c}
\big(D^-\cdot 1\!\!{\rm I}_{2_c}\big)Q^c.\label{IV.8}
\end{eqnarray}
In the following, it is very convenient to use the Nambu--Gorkov
formalism, in which $Q$ quarks are composed into a bispinor $\Psi$ such
that
\begin{equation}
\Psi=\left({Q\atop Q^c}\right),~~\Psi^T=(Q^T,\overline Q C^T);~~
\quad \overline\Psi=(\overline Q,\overline { Q^c})=(\overline Q,Q^T C)=\Psi^T \left
(\begin{array}{cc}
0~~,&  C\\
C~~, &0
\end{array}\right )\equiv\Psi^T Y.
\label{140}
\end{equation}
Now, taking into account the Eq. (\ref{IV.8}) and introducing the matrix-valued operator $Z$ (see in Eq. (\ref{15})),
one can rewrite the functional Gaussian integral over $Q$ and $\overline Q$ in (\ref{21}) in
terms of $\Psi$ and $Z$ and then evaluate it as follows
(clearly, in this case $[d\overline Q][dQ]=$ $[dQ^c][dQ]=$ $[d\Psi]$):
\begin{eqnarray}
&&\int [d\overline Q][dQ]\exp\Bigl(i\int\Big [\overline Q \big(D^+\cdot 1\!\!{\rm I}_{2_c}\big)Q-
 \frac{\Delta}{2}[\overline{Q^c}i\gamma^5\tau_2\sigma_{2} Q]
-\frac{\Delta}{2}[\overline Q i\gamma^5\tau_2\sigma_{2}Q^c]\Big ]d^4 x\Bigr)\nonumber\\
&=&  \int[d\Psi]\exp\left\{\frac
i2\int\overline\Psi Z\Psi d^4x\right\}=
  \int[d\Psi]\exp\left\{\frac i2\int\Psi^T(YZ)\Psi
  d^4x\right\}=\mbox {det}^{1/2}(YZ)=\mbox {det}^{1/2}(Z),
\label{210}
\end{eqnarray}
where the last equality is valid due to the evident relation $\det Y=1$. Now, using a general formula
\begin{eqnarray}
\det\left
(\begin{array}{cc}
A~, & B\\
C~, & D
\end{array}\right )=\det [-CB+CAC^{-1}D]=\det
[DA-DBD^{-1}C],
\label{IV.21}
\end{eqnarray}
it is possible to find that (the notations used below are introduced in Eqs. (\ref{15}) and (\ref{14}))
\begin{eqnarray}
&&\det(Z)=\det\left
(\begin{array}{cc}
D^+\cdot 1\!\!{\rm I}_{2_c}, & -K\\
~-K~~~~, & D^-\cdot 1\!\!{\rm I}_{2_c}
\end{array}\right )=\det [\big(-KK+KD^+K^{-1}D^-\big)\cdot 1\!\!{\rm I}_{2_c}]\label{IV.220}\\
&&=\det \Big [\Delta^2\cdot 1\!\!{\rm I}_{2_c}+\big (-i\hat\partial-m-\widetilde{\cal M}\gamma^0 -\sigma +i\gamma^5\tau_1\pi_1\big )
\big (i\hat\partial-m-\gamma^0{\cal M} -\sigma -i\gamma^5\tau_1\pi_1\big )\cdot 1\!\!{\rm I}_{2_c}\Big ],\label{IV.22}
\end{eqnarray}
where we use the notations $\hat\partial=\gamma^\alpha\partial_\alpha$, $\mu=\mu_B/3, \nu=\mu_I/2,\nu_5=\mu_{I5}/2$ and
\begin{eqnarray}
\widetilde{\cal M}=\mu+\mu_5\gamma^5-\nu\tau_3-\nu_{5}\gamma^5\tau_3. \label{IV.23}\end{eqnarray}
 
In the particular case when $\Delta=0$ (in this case the diquark channel of the NJL model (1) is ignored) it is clear from Eq. (\ref{IV.220}) that 
\begin{eqnarray}
&&\det(Z)=\det [\gamma^5\tau_2D^+\gamma^5\tau_2D^-\cdot 1\!\!{\rm I}_{2_c}]={\rm \det}^2 [D^+D^-]={\rm \det}^4 (D^+).\label{IV.221}
\end{eqnarray}
The last equality in this exprerssion follows from Eq. (\ref{IV.70}), which means that  $\det(D^-)=\det(D^+)$. Taking into account in Eq. 
(\ref{16}) the relation (\ref{IV.221}), we see that at $\Delta=0$ the quark contribution to the effective action (\ref{16}) is equal to 
$-3i\ln\det(D^+)$, i.e. coinsides with a corresponding expression for the effective action of the NJL model (1) with $N_c=3$ and $H=0$
(see in Ref. \cite{kkz18}).

Since $\hat\partial\hat\partial=\partial^2$, we have from Eq. (\ref{IV.22})
\begin{eqnarray}
&&\det(Z)=\det \Big [\big(\Delta^2+\partial^2+i\hat\partial {\cal A} +{\cal B}i\hat\partial  -{\cal B}{\cal A}\big)\cdot 1\!\!{\rm I}_{2_c}
\Big ],\label{IV.24}
\end{eqnarray}
where 
\begin{eqnarray}
{\cal A}=m+\sigma+\gamma^0{\cal M}+i\gamma^5\tau_1\pi_1,~~{\cal B}=-m-\sigma-\widetilde{\cal M}\gamma^0+
i\gamma^5\tau_1\pi_1.\label{IV.25}
\end{eqnarray}
Remark that the expression in the square brackets of Eq. (\ref{IV.24}) is proportional to a unit 
operator in the two-dimentional (i.e. $N_c=2$ below) color space, so it follows from Eq. (\ref{IV.24}) that
\begin{eqnarray}
&&\det(Z)\equiv{\rm det}^{N_c}{\cal D}={\rm det}^{N_c}\left
(\begin{array}{cc}
D_{11}~, & D_{12}\\
D_{21}~, & D_{22}
\end{array}\right ),\label{IV.240}
\end{eqnarray}
where ${\cal D}$ is the $2\times 2$ matrix in the 2-dim flavor space (its matrix elements $D_{kl}$ are the nontrivial operators in
the 4-dim spinor and in the (3+1)-dim coordinate spaces). To get the exact expressions of the matrix elements $D_{kl}$, we need to bear 
in mind that
\begin{eqnarray}
i\hat\partial {\cal A} &+&{\cal B}i\hat\partial=i\hat\partial \gamma^0{\cal M}-\widetilde{\cal M}\gamma^0i\hat\partial=
(\mu+\mu_5\gamma^5)\big[i\hat\partial \gamma^0-\gamma^0i\hat\partial \big ]+(\nu+\nu_5\gamma^5)\big[i\hat\partial \gamma^0+\gamma^0i\hat\partial
\big ]\tau_3,\nonumber\\
-{\cal B}{\cal A}&=&\pi_1^2+(m+\sigma+\widetilde{\cal M}\gamma^0)(m+\sigma+\gamma^0{\cal M})-i\gamma^5\tau_1\pi_1
\gamma^0{\cal M}+\widetilde{\cal M}\gamma^0i\gamma^5\tau_1\pi_1\nonumber\\
&&\hspace{-1cm}=\pi_1^2+(m+\sigma)^2+2(m+\sigma)\gamma^0(\mu+\nu_5\gamma^5\tau_3)+\widetilde{\cal M}{\cal M}-i\gamma^5\tau_1\pi_1
\gamma^0{\cal M}+\widetilde{\cal M}\gamma^0i\gamma^5\tau_1\pi_1,\label{IV.26}
\end{eqnarray}
Note also that
\begin{eqnarray}
\widetilde{\cal M}{\cal M}&=&(\mu+\mu_5\gamma^5)^2-(\nu+\nu_5\gamma^5)^2,\nonumber\\
-i\gamma^5\tau_1\pi_1\gamma^0{\cal M}&=&i\big (\mu-\mu_5\gamma^5-\nu\tau_3+\nu_5\gamma^5\tau_3\big )\gamma^0\gamma^5
\tau_1\pi_1,\nonumber\\
i\widetilde{\cal M}\gamma^0\gamma^5\tau_1\pi_1&=&i\big (\mu+\mu_5\gamma^5-\nu\tau_3-\nu_5\gamma^5\tau_3\big )\gamma^0\gamma^5
\tau_1\pi_1,\nonumber\\
-i\gamma^5\tau_1\pi_1\gamma^0{\cal M}&+&i\widetilde{\cal M}\gamma^0\gamma^5\tau_1\pi_1=
2i\big (\mu-\nu\tau_3\big )\gamma^0\gamma^5\tau_1\pi_1
=2i\mu\gamma^0\gamma^5\tau_1\pi_1+2\nu\gamma^0\gamma^5\tau_2\pi_1.\label{IV.30}
\end{eqnarray}
Now, taking into account the relations (\ref{IV.24})-(\ref{IV.30})
we are ready to present the expressions for the matrix elements $D_{kl}$ of the 2$\times$2 matrix ${\cal D}$ from Eq. (\ref{IV.240}),
\begin{eqnarray}
D_{11}&=&\Delta^2+\partial^2+(\mu+\mu_5\gamma^5)\big[i\hat\partial \gamma^0-\gamma^0i\hat\partial \big ]+
(\nu+\nu_5\gamma^5)\big[i\hat\partial \gamma^0+\gamma^0i\hat\partial
\big ]+\pi_1^2+(m+\sigma)^2\nonumber\\
&+&2(m+\sigma)\gamma^0(\mu+\nu_5\gamma^5)+(\mu+\mu_5\gamma^5)^2-(\nu+\nu_5\gamma^5)^2,\nonumber\\
D_{22}&=&\Delta^2+\partial^2+(\mu+\mu_5\gamma^5)\big[i\hat\partial \gamma^0-\gamma^0i\hat\partial \big ]
-(\nu+\nu_5\gamma^5)\big[i\hat\partial \gamma^0+\gamma^0i\hat\partial
\big ]+\pi_1^2+(m+\sigma)^2\nonumber\\
&+&2(m+\sigma)\gamma^0(\mu-\nu_5\gamma^5)+(\mu+\mu_5\gamma^5)^2-(\nu+\nu_5\gamma^5)^2,\nonumber\\
D_{12}&=&2i\gamma^0\gamma^5(\mu-\nu)\pi_1,~~~D_{21}=2i\gamma^0\gamma^5(\nu+\mu)\pi_1.
\label{IV.31}
\end{eqnarray}
Due to a rather general formula $\det O=\exp {\rm Tr}\ln O$, we see from Eq. (\ref{16}) that indeed, in order to calculate an effective action,
we should evaluate only the quantity ${\rm Tr}\ln Z$. Then, taking into account the general relations (\ref{B5}) and (\ref{B6}) as well as Eq. 
(\ref{IV.240}), one can obtain
\begin{eqnarray}
{\rm Tr}\ln Z &=& N_c\ln\det {\cal D}=N_c{\rm Tr}\ln{\cal D}=N_c\int\frac{d^4p}{(2\pi)^4}{\rm tr}\ln\overline{\cal D}(p)\int d^4x\nonumber\\
&=&N_c\int\frac{d^4p}{(2\pi)^4}\ln\det\overline{\cal D}(p)\int d^4x,
\label{IV.32}
\end{eqnarray}
where ${\rm Tr}$ means the trace of an operator both in the coordinate and internal spaces, whereas {\rm tr} is the trace only in internal space. Moreover,
$\overline{\cal D}(p)$ is the 2$\times$2 matrix which is the momentum space representation of the matrix ${\cal D}$ from Eq. (\ref{IV.240}). Its matrix elements
$\overline{\cal D}_{kl}(p)$ can be obtained from the relations (\ref{IV.31}) by simple replacements, $i\hat\partial\to\hat p$ and $\partial^2\to-p^2$. So we
have from (\ref{IV.31})
\begin{eqnarray}
\overline{D}_{11}(p)&=&\Delta^2-p^2+(\mu+\mu_5\gamma^5)\big[\hat p \gamma^0-\gamma^0\hat p \big ]+
(\nu+\nu_5\gamma^5)\big[\hat p \gamma^0+\gamma^0\hat p\big ]+\pi_1^2+(m+\sigma)^2\nonumber\\
&+&2(m+\sigma)\gamma^0(\mu+\nu_5\gamma^5)+(\mu+\mu_5\gamma^5)^2-(\nu+\nu_5\gamma^5)^2,\nonumber\\
\overline{D}_{22}(p)&=&\Delta^2-p^2+(\mu+\mu_5\gamma^5)\big[\hat p \gamma^0-\gamma^0\hat p \big ]
-(\nu+\nu_5\gamma^5)\big[\hat p\gamma^0+\gamma^0\hat p\big ]+\pi_1^2+(m+\sigma)^2\nonumber\\
&+&2(m+\sigma)\gamma^0(\mu-\nu_5\gamma^5)+(\mu+\mu_5\gamma^5)^2-(\nu+\nu_5\gamma^5)^2,\nonumber\\
\overline{D}_{12}(p)&=&2i\gamma^0\gamma^5(\mu-\nu)\pi_1,~~~
\overline{D}_{21}(p)=2i\gamma^0\gamma^5(\nu+\mu)\pi_1.
\label{IV.33}
\end{eqnarray}
In Eq. (\ref{IV.32}) we should evaluate the determinant of the operator $\overline{\cal D}(p)$ which is a 2$\times$2 matrix in flavor space and 4$\times$4 matrix
in spinor space. It is a rather difficult task. But due to a general relation (\ref{IV.21}), we have
\begin{eqnarray}
&&\det\overline{D}(p)\equiv {\rm det}\left
(\begin{array}{cc}
\overline{D}_{11}(p)~, & \overline{D}_{12}(p)\\
\overline{D}_{21}(p)~, & \overline{D}_{22}(p)
\end{array}\right )\nonumber\\
&=&\det\Big[-\overline{D}_{21}(p)\overline{D}_{12}(p)+\overline{D}_{21}(p)\overline{D}_{11}(p)\big(\overline{D}_{21}(p)\big)^{-1}
\overline{D}_{22}(p)\Big]\equiv\det L(p),\label{IV.34}
\end{eqnarray}
where the matrix $L(p)$, i.e. the expression/matrix in square brackets of Eq. (\ref{IV.34}), is indeed a 4$\times$4 matrix in 4-dim
spinor space only, which is composed of 4$\times$4 matrices $\overline{D}_{ij}(p)$ (see in Eq. (\ref{IV.33})). So, since indeed $N_c=2$,
\begin{eqnarray}
\ln\det Z={\rm Tr}\ln Z =2\int\frac{d^4p}{(2\pi)^4}\ln\det L(p)\int d^4x.
\label{IV.35}
\end{eqnarray}

\section{Traces of operators and their products}
\label{ApB}

Let $\hat A,\hat B,...$ are some operators in the Hilbert space
$\mathbf H$ of functions $f(x)$ depending on four real variables,
$x\equiv (x^0,x^1,x^2,x^3)$. In the coordinate representation their matrix
elements are $A(x,y), B(x,y) ,...$, correspondingly, so that
 $$(\hat A f)(x)\equiv \int d^4yA(x,y)f(y),~~(\hat A\cdot \hat
 B)(x,y)\equiv \int
 d^4zA(x,z)B(z,y), ~~\mbox{etc}$$
 By definition,
 \begin{eqnarray}
{\rm Tr}\hat A\equiv\int d^4xA(x,x),~~{\rm Tr}(\hat A\cdot\hat
B)\equiv\int d^4xd^4yA(x,y)B(y,x),~~\mbox{etc}.
\label{B0}
\end{eqnarray}
Now suppose that $A(x,y)\equiv A(x-y)$, $B(x,y)\equiv B(x-y)$, i.e.
that $\hat A,\hat B$ are translationally invariant operators. Then
introducing the Fourier transformations of their matrix elements,
i.e.
\begin{eqnarray}
\overline{A}(p)=\int d^4z
A(z)e^{ipz},
~~~~~~~ A(z)=\int\frac{d^4p}{(2\pi)^4}
\overline{A}(p)e^{-ipz},~~~\mbox{etc},
\label{B3}
\end{eqnarray}
where $z=x-y$, it is possible to obtain from the above formulae
 \begin{eqnarray}
{\rm Tr}\hat A=A(0)\int d^4x=\int\frac{d^4p}{(2\pi)^4}
\overline{A}(p)\int d^4x.
\label{B4}
\end{eqnarray}
If there is an operator function $F(\hat A)$, where $\hat A$ is a
translationally invariant operator, then in the coordinate
representation its matrix elements depend on the difference $(x-y)$.
Obviously, it is possible to define the Fourier transformations
$\overline{F(A)}(p)$ of its matrix elements, and the following
relations are valid ($\overline{A}(p)$ is the Fourier transformation (\ref{B3})
for the matrix element $A(x-y)$):
\begin{eqnarray}
\overline{F(A)}(p)=F(\overline{A}(p));~~~
{\rm Tr}F(\hat A)=\int\frac{d^4p}{(2\pi)^4}
F(\overline{A}(p))\int d^4x.
\label{B5}
\end{eqnarray}
Finally, suppose that $\hat A$ is an operator in some internal
$n$-dimensional vector space, in addition. Evidently, the same is
valid for the Fourier transformation $\overline{A}(p)$ which is now
some $n\times n$ matrix. Let $\lambda_i(p)$ are eigenvalues of the
$n\times n$ matrix $\overline{A}(p)$, where  $i=1,2,..,n$. Then
\begin{eqnarray}
{\rm Tr}F(\hat A)=\int\frac{d^4p}{(2\pi)^4}
\mbox{tr}F(\overline{A}(p))\int
d^4x=\sum_{i=1}^{n}\int\frac{d^4p}{(2\pi)^4}
F(\lambda_i(p))\int d^4x.
\label{B6}
\end{eqnarray}
In this formula we use the notation tr for the trace of any operator
in the internal $n$-dimensional vector space only, whereas the
symbol Tr means the trace of an operator both in the coordinate and
internal spaces. In particular, if $F(\hat A)=\ln(\hat A)$, then it follows from (\ref{B6}) that (here we use a well-known relation 
$\ln\det (\hat A)={\rm Tr}\ln (\hat A)$)
\begin{eqnarray}
\ln\det (\hat A)={\rm Tr}\ln (\hat A)=\sum_{i=1}^{n}\int\frac{d^4p}{(2\pi)^4}
\ln(\lambda_i(p))\int d^4x=
\int\frac{d^4p}{(2\pi)^4}
\ln(\det\overline{A}(p))\int d^4x.
\label{B7}
\end{eqnarray}


\begin{thebibliography}{999}

\bibitem{njl}
Y. Nambu and G. Jona-Lasinio, Phys. Rev. {\bf 122}, 345 (1961); {\bf 124}, 246 (1961).

\bibitem{buballa2}
S.~P.~Klevansky,
  Rev.\ Mod.\ Phys.\  {\bf 64}, 649 (1992);
D. Ebert, H. Reinhardt and M. K. Volkov, Prog. Part. Nucl. Phys. {\bf 33}, 1 (1994); 
T.~Inagaki, T.~Muta and S.~D.~Odintsov,
  Prog.\ Theor.\ Phys.\ Suppl.\  {\bf 127}, 93 (1997);
A. A. Garibli, R. G. Jafarov, and V. E. Rochev, 
Symmetry {\bf 11}, no. 5, 668 (2019).

\bibitem{buballa}
M. Buballa, Phys. Rep. {\bf 407}, 205 (2005).

\bibitem{zhokhov}
T.~G.~Khunjua, K.~G.~Klimenko and R.~N.~Zhokhov,
 Symmetry {\bf 11}, no. 6, 778 (2019);
  Particles {\bf 3}, no. 1, 62 (2020).
  
\bibitem{son}
D. T.~Son and M. A.~Stephanov, Phys.\ Atom.\ Nucl.\  {\bf 64}, 834 (2001);
M. Loewe and C. Villavicencio, Phys. Rev. D {\bf 67}, 074034
(2003); M.~Frank, M.~Buballa and M.~Oertel,
Phys. Lett. B \textbf{562}, 221-226 (2003);
D. C.~Duarte, R. L. S.~Farias and R. O.~Ramos,
  Phys.\ Rev.\  D {\bf 84}, 083525 (2011);
D.~Ebert, K. G.~Klimenko, A. V.~Tyukov and V. C.~Zhukovsky,
  Eur.\ Phys.\ J.\ C {\bf 58}, 57 (2008).

\bibitem{he}
L. He, M. Jin, and P. Zhuang, Phys. Rev. D {\bf 71}, 116001 (2005);
D. Ebert and K. G. Klimenko, J.\ Phys.\ G {\bf 32}, 599 (2006);
Eur.\ Phys.\ J.\  C {\bf 46}, 771 (2006);
C.f.~Mu, L.y.~He and Y.x.~Liu,
  Phys.\ Rev.\  D {\bf 82}, 056006 (2010).

\bibitem{ak}
J. O.~Andersen and T.~Brauner,
  Phys.\ Rev.\  D {\bf 78}, 014030 (2008);
J. O.~Andersen and L.~Kyllingstad,
 J.\ Phys.\ G {\bf 37}, 015003 (2009);
 Y.~Jiang, K.~Ren, T.~Xia and P.~Zhuang,
  Eur.\ Phys.\ J.\ C {\bf 71}, 1822 (2011);
  A.~Folkestad and J.~O.~Andersen,
  Phys.\ Rev.\ D {\bf 99}, 054006 (2019);
   P.~Adhikari, J.~O.~Andersen and P.~Kneschke,
   Phys.\ Rev.\ D {\bf 98},  074016  (2018);
   Eur.\ Phys.\ J.\ C {\bf 79},  874 (2019);
  J.~O.~Andersen, P.~Adhikari and P.~Kneschke,
PoS \textbf{Confinement2018}, 197 (2019) [arXiv:1810.00419 [hep-ph]].

\bibitem{ekkz}
 D.~Ebert, T. G.~Khunjua, K. G.~Klimenko and V. C.~Zhukovsky,
  Int.\ J.\ Mod.\ Phys.\ A {\bf 27}, 1250162 (2012);
  N. V.~Gubina, K. G.~Klimenko, S. G.~Kurbanov and V. C.~Zhukovsky,
  Phys.\ Rev.\ D {\bf 86}, 085011 (2012).

  \bibitem{Mammarella:2015pxa} 
  A.~Mammarella and M.~Mannarelli,
  Phys.\ Rev.\ D {\bf 92},  085025 (2015);
S.~Carignano, L.~Lepori, A.~Mammarella, M.~Mannarelli and G.~Pagliaroli,
  Eur.\ Phys.\ J.\ A {\bf 53},  35 (2017);
M.~Mannarelli,
  Particles {\bf 2}, no. 3, 411 (2019).
  
\bibitem{Andersen:2018nzq}
       J.~O.~Andersen and P.~Kneschke,
       arXiv:1807.08951 [hep-ph];
       B.~B.~Brandt, G.~Endrodi, E.~S.~Fraga, M.~Hippert, J.~Schaffner-Bielich and S.~Schmalzbauer,
       Phys.\ Rev.\ D {\bf 98},  094510 (2018).

\bibitem{Khunjua:2019lbv}
T.~G.~Khunjua, K.~G.~Klimenko and R.~N.~Zhokhov,
JHEP \textbf{06}, 006 (2019).

\bibitem{kkz18}
T.~G.~Khunjua, K.~G.~Klimenko and R.~N.~Zhokhov,
Phys. Rev. D \textbf{98}, no.5, 054030 (2018).

\bibitem{kkz18-2}
T.~G.~Khunjua, K.~G.~Klimenko and R.~N.~Zhokhov,
Phys. Rev. D \textbf{97}, no.5, 054036 (2018).

\bibitem{khunjua}
T.~G.~Khunjua, K.~G.~Klimenko and R.~N.~Zhokhov,
Phys. Rev. D \textbf{100}, no.3, 034009 (2019); Moscow Univ. Phys. Bull. \textbf{74}, no.5, 473 (2019); Eur. Phys. J. C \textbf{80}, no.10, 995 (2020); 
Acta Phys. Polon. Supp. \textbf{14}, 67 (2021).

\bibitem{khunjua2}
T.~G.~Khunjua, K.~G.~Klimenko and R.~N.~Zhokhov,
Eur. Phys. J. C \textbf{79}, no.2, 151 (2019).


\bibitem{alford}
I. A. Shovkovy, Found. Phys. {\bf 35}, 1309 (2005);
 M.~Huang,
  Int.\ J.\ Mod.\ Phys.\ E {\bf 14}, 675 (2005);
 K.~G.~Klimenko and D.~Ebert,
  Theor.\ Math.\ Phys.\  {\bf 150}, 82 (2007) [Teor.\ Mat.\ Fiz.\  {\bf 150}, 95 (2007)];
 M. G.~Alford, A.~Schmitt, K.~Rajagopal, and T.~Sch\"afer,
 Rev.\ Mod.\ Phys.\  {\bf 80}, 1455 (2008);
 E.~J.~Ferrer and V.~de la Incera,
  Lect.\ Notes Phys.\  {\bf 871}, 399 (2013).

\bibitem{Ebert}
D.~Ebert, K.~G.~Klimenko, V.~C.~Zhukovsky and A.~M.~Fedotov,
Eur. Phys. J. C \textbf{49}, 709 (2007).

\bibitem{2color}
T.~G.~Khunjua, K.~G.~Klimenko and R.~N.~Zhokhov,
Phys. Rev. D \textbf{106}, no.4, 045008 (2022).

\end{thebibliography}
\end{document}